\documentclass[letterpaper, 10 pt, conference]{ieeeconf}  
\IEEEoverridecommandlockouts                     
\usepackage{amssymb}  
\usepackage{amsmath}
\usepackage{cite}
\usepackage{booktabs}
\usepackage{pdfpages}
\usepackage{graphicx}
\usepackage{lipsum}
\usepackage{mathtools}
\usepackage{cuted}
\usepackage[T1]{fontenc}
\usepackage[utf8]{luainputenc}
\usepackage{amsbsy}
\usepackage{algorithm}
\usepackage{algpseudocode}
\usepackage{algpascal}
\usepackage{bm}
\usepackage{hyperref}
\usepackage[font=small,skip=0pt]{caption}
               
\usepackage{siunitx}

\usepackage{colortbl, array,xcolor}
\definecolor{myblue}{rgb}{0.5647,0.6784,0.8824}

\newcommand{\myTableCell}[1]{%
  {\renewcommand{\arraystretch}{1.0}%
  \begin{tabular}[c]{@{}l@{}}#1\end{tabular}}%
}

\usepackage{cite}
\usepackage{epsfig}

\title{\LARGE \bf
Machine-knittable, Magnetically-Plug-n-Play E-Textile Prototyping
}

\author{ Yifan Li*$^{1}$, Ryo Takahashi*$^{1}$, Wakako Yukita$^{1}$, Irmandy Wicaksono$^{2}$, Kanata Matsutani$^{3}$, Yuhiro Iwamoto$^{3}$, \\ Sunghoon Lee$^{1}$, Tomoyuki Yokota$^{1}$, Takao Someya$^{1}$, Yoshihiro Kawahara$^{1}$ 
\thanks{This work was mainly supported by JST JPMJCS25N4, JPMJPR2515, JPMJAP2401, JST SPRING JPMJSP2108 and JSPS 22K21343. \it(*co-first authors, $\dagger$Corresponding author: Ryo Takahashi)}
\thanks{$^{1}$Y. Li, R. Takahashi, W. Yukita, T. Yokota, T. Someya, Y. Kawahara are with Graduate School of Engineering, The University of Tokyo, Tokyo, Japan.
        {\tt\small takahashi@akg.t.u-tokyo.ac.jp}}%
\thanks{$^{2}$I, Wicaksono is with the Division of Industrial Design, National University of Singapore, Singapore.
        {\tt\small irmandy@nus.edu.sg}}%
\thanks{$^{3}$ K. Matsutani and Y. Iwamoto are Graduate School of Engineering Nagare College, Nagoya Institute of Technology, Nagoya, Japan.
        {\tt\small iwamoto.yuhiro@nitech.ac.jp}
}%
}

\begin{document}

\maketitle

\begin{abstract}
Electronic textiles (e-textiles) integrated with wearable sensors are essential for daily motion monitoring and long-term physiological sensing.
For example, capturing optimal kinematic or bio-signals requires aligning sensors with specific anatomical parts, which vary significantly across individuals and application scenarios.
This necessity for personalization makes e-textile prototyping inherently iterative, however current fabrication methods, such as manual conductive stitching, rely on permanent bonds that restrict rapid adjustment.
This paper introduces \textit{Plug-n-play e-knit}, a machine-knittable e-textile prototyping platform that enables repeatable, quick adjustment of sensor positions across garments.
First, to cover the large area of the textile for prototyping, we use industrial digital knitting of conductive yarn to integrate power and communication buses directly into the large-scale textile. 
Then, to ensure plug-n-play attachment to the textile, we employ soft-magnetic connectors that enable sensors to be repeatedly plugged into the wiring without damaging the fabric.
Furthermore, our LED-positioning system enables the automatic identification and localization of each sensor node.
We demonstrate the platform's capabilities through forearm movement calibration and position-aware temperature mapping.

\end{abstract}

\section{INTRODUCTION}

Electronic textiles (e-textiles) offer a promising approach for continuous monitoring of the human body without bulky equipment~\cite{ohiri_e-textile_2022, spHRI2025}. 
To capture high-quality data, such as the exact angle of a knee bend, sensors must be placed precisely on specific anatomical spots~\cite{hill_threadboard_2021}. 
However, the human body is not a one-size-fits-all platform. 
The ideal sensor location varies significantly across individuals and even within the same individual, depending on the activity~\cite{woelfle_plug-and-play_2023}.
Conventional e-textiles cannot easily adapt to these changes. 
Most e-textiles use pre-designed power and data networks that are hand-stitched into the fabric with conductive threads to ensure a stable connection~\cite{fromme_metal-textile_2021,takahashi2021meander}.
If a researcher or user needs to move a sensor by just a few centimeters to improve the signal, they often have to tear the threads, which damages the garment~\cite{stanley_modular_2022}.
While some modular designs try to solve this using metal snap buttons~\cite{jones_swatch-bits_2020} or clips~\cite{LilyPad,Lui_Reconstruction_Kit}, these add hard spots to the soft fabric, making the clothes uncomfortable to wear and prone to breaking when the fabric stretches~\cite{liu2025sensory_review, azani2024electronic}.
Furthermore, these designs cannot accommodate significant shifts in sensor placement, because the electrical networks are typically restricted to specific one-dimensional paths~\cite{lin_digitally-embroidered_2022}.

\begin{figure*}[h!]
  \centering
  \includegraphics[width=1.0\textwidth]{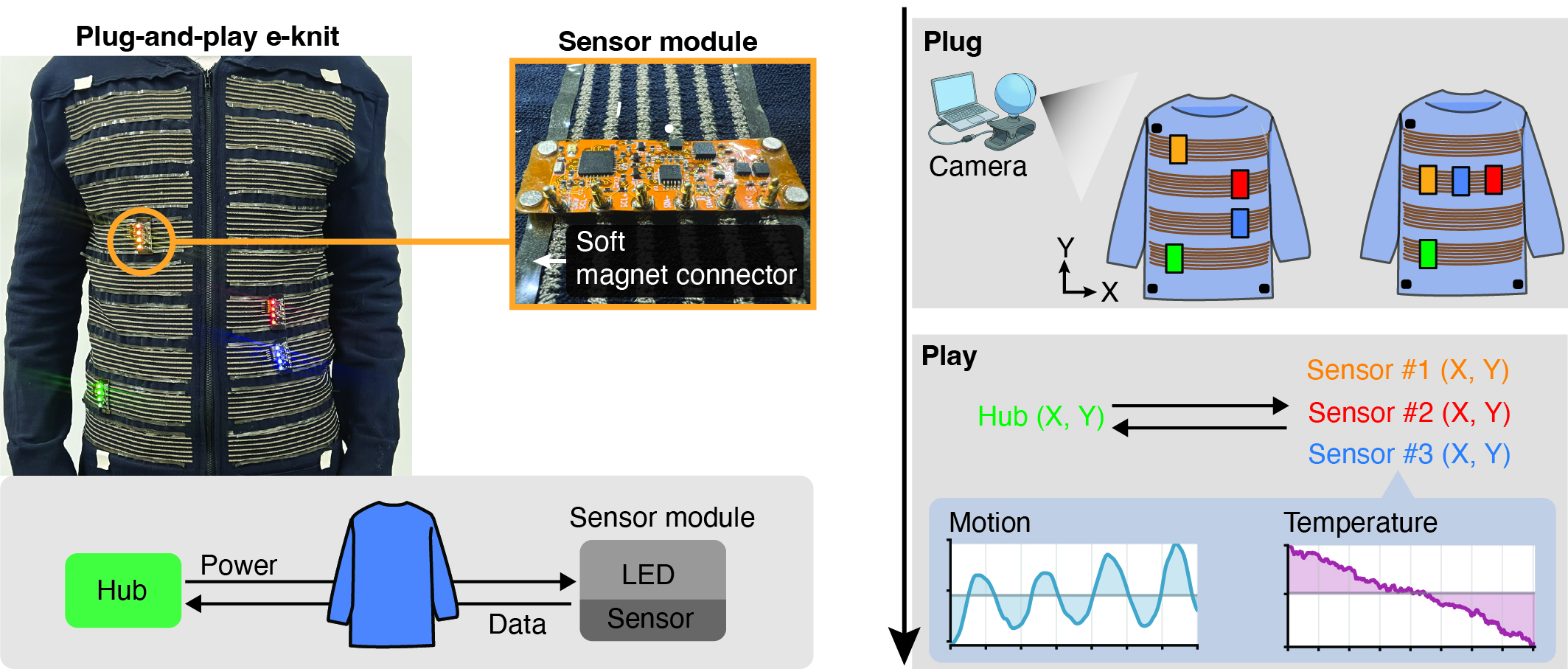}
  \caption{Design overview of plug-n-play e-knit comprising the textile garment, sensing module, soft magnet and control hub. When the user places the sensor module on the textile in a plug-n-play way, a hub will power the LEDs on the sensor modules, while an external camera can position all sensor modules' IDs and positions. Then, the modules start to sense data and communicate with the hub.}
  \label{fig:Overview}
\end{figure*}

To bridge this gap, we introduce \textit{Plug-n-play e-knit}, a large-scale e-textile platform designed for rapid and non-destructive reconfiguration (see Figure~\ref{fig:Overview}). 
Our system transforms a garment into a smart canvas where sensors can be placed and moved easily.
The platform consists of two main technologies. 
First, to embed power and data lines into the large-scale textiles without a manual process, we use industrial digital knitting of conductive yarns, enabling automatic integration of the textile-based wiring. 
Second, we developed soft magnetic connectors, allowing sensing modules to snap onto the integrated wires, allowing sensors to be repositioned hundreds of times without any damage.
Together, plug-n-play e-knit platform provides a robust framework for anyone to design, test, and optimize wearable technology in minutes.
Unlike previous plug-n-play works (~\autoref{tab:comparison}) limited by manual fabrication or low node capacities, our platform achieves automated large-scale coverage ($>$70\%) and supports $>$20 sensors. While the similar I$^2$We~\cite{noda_I2We_2019} uses a double-layer woven structure with rigid sockets for widespread distribution, its large overlapping fabrics introduce significant stray capacitance, slowing signal transitions and restricting maximum sensor capacity. 
In contrast, our method leverages differential I$^2$C and low-resistance conductive threads to reject electrical noise, seamlessly supporting a dense sensor network.
Our proof-of-concept plug-n-play e-knit shows two scenarios of e-textile prototyping, including i) motion capturing test for finding the optimal sensor placement and ii) on-textile temperature recording in a room.

\begin{table*}[h!]
    \renewcommand{\arraystretch}{1.4}
    \centering
    \caption{Technical comparison of plug-n-play e-textile.}
    \begin{tabular}{l|llllll}\toprule
        \textbf{E-textile} & 
        \textbf{\myTableCell{Fabrication}} & \textbf{\myTableCell{Connection \\method}} & 
        \textbf{\myTableCell{Damage to\\textile}} &
        \textbf{\myTableCell{Communication\\method}} &
        \textbf{\myTableCell{Coverage\\(torso)}} &
        \textbf{\myTableCell{Available \\sensor number}}\\\hline\hline
        \rowcolor{myblue!20}Modular E-textile~\cite{stanley_modular_2022}& Manual stitching & \myTableCell{Custom stretchable\\connector} & \pmb{\checkmark} & I$^2$C & <40\% & <10 \\ 
        \myTableCell{Plug-n-play Wearables~\cite{woelfle_plug-and-play_2023}} & Manual stitching & LumeoLoop~\cite{LumeoLoop} & \pmb{\checkmark} & - & <20\% & <5 \\ 
        \rowcolor{myblue!20}I$^2$We~\cite{noda_I2We_2019}& Machine-weaving & Socket & \pmb{\checkmark} & I$^2$C & >70\% & <10 \\ 
        \textbf{Plug-n-play e-knit}~& Machine-knitting & \myTableCell{Soft magnet} & $\times$ & Differential I$^2$C & >70\% & >20 \\
        \bottomrule
    \end{tabular}
    \label{tab:comparison}
\end{table*}

\section{Plug-n-Play E-knit Design}

Our system integrates a machine-knitted electronic textile with detachable sensing modules to create a robust, reconfigurable wearable platform. 
The textile features a standardized infrastructure of six horizontal conductive threads that serve as independent power and data channels. 
The outermost threads provide a stable 5V power supply ($V_{\rm CC}$) and ground (GND), while the inner four channels handle data communication. 
We adopted the Inter-Integrated Circuit (I$^2$C) protocol because its shared-bus architecture naturally supports connecting multiple addressable sensors for plug-n-play scalability~\cite{Righetti_I2C}.
To overcome the signal noise and limited range typical of fabric-based circuits, differential I$^2$C is implemented. 
This protocol transmits signals as differential pairs, doubling the effective amplitude at the receiver and canceling electromagnetic interference.
This ensures reliable, high-speed data transmission across the garment’s surface, even during body movement.

The system achieves a seamless plug-n-play experience through a combination of magnetic self-alignment and automated vision tracking. 
We utilized permanent Magnet Elastomer (PME)---soft, strip-shaped magnets---affixed to the textile to act as an interface for the sensor modules. When a module is placed on the fabric, rigid magnets on its corners snap to the PME, automatically compressing spring-loaded pogo pins against the conductive threads to establish a secure electrical link. 
Simultaneously, an external camera system identifies the module’s location and identity. 
By detecting corner markers to define the textile's boundaries and analyzing the specific HSV color space of a module’s onboard LED, the system automatically maps each sensor's precise coordinates and ID. 
This allows for rapid network reconfiguration without the need for manual sewing or complex software setup.

\section{Implementation}

\subsection{E-textile Fabrication}

\autoref{fig:PME} illustrates the prototype of the
plug-n-play e-knit. The fabrication process is as follows: First, the textile pattern was designed and knitted using an industrial knitting machine (MACH2XS, SHIMA SEIKI), as shown in \autoref{fig:PME}a. 
We selected machine-knitting over weaving because its continuous looping process is more suited for e-textile production, allowing complex conductive patterns and insulating layers in a single, automated step~\cite{kalicaktowards}.
The e-textile integrates a total of six differential I$^2$C line groups circling the torso, three across the chest, two on the back, and one connecting both arms and shoulders~\autoref{fig:PME}b. 
To embed the differential I$^2$C signal wiring into the knitted textile, the two-layer jacquard knitting technique was employed, which ensures physical insulation of the wiring against the human skin. 
Each differential I²C line group (ground line, power supply line, two pairs of differential SCL lines, and SDA lines) consists of 2 mm-wide conductive threads (AGposs, Mitsufuji Corporation), with a 4 mm spacing between them. 
To interconnect these horizontal lines, we employed a vertical textile strip of the same construction, overlaying it on the back of the textile to bridge the lines.
In total, the wiring covers over approximately 4680 cm$^2$ around the torso, and extends 150.5 cm on arms with a resistance of 102.3 $\Omega$. 

\subsection{Soft Magnet Connector}


\begin{figure*}[h]
  \centering
  \includegraphics[width=1.0\textwidth]{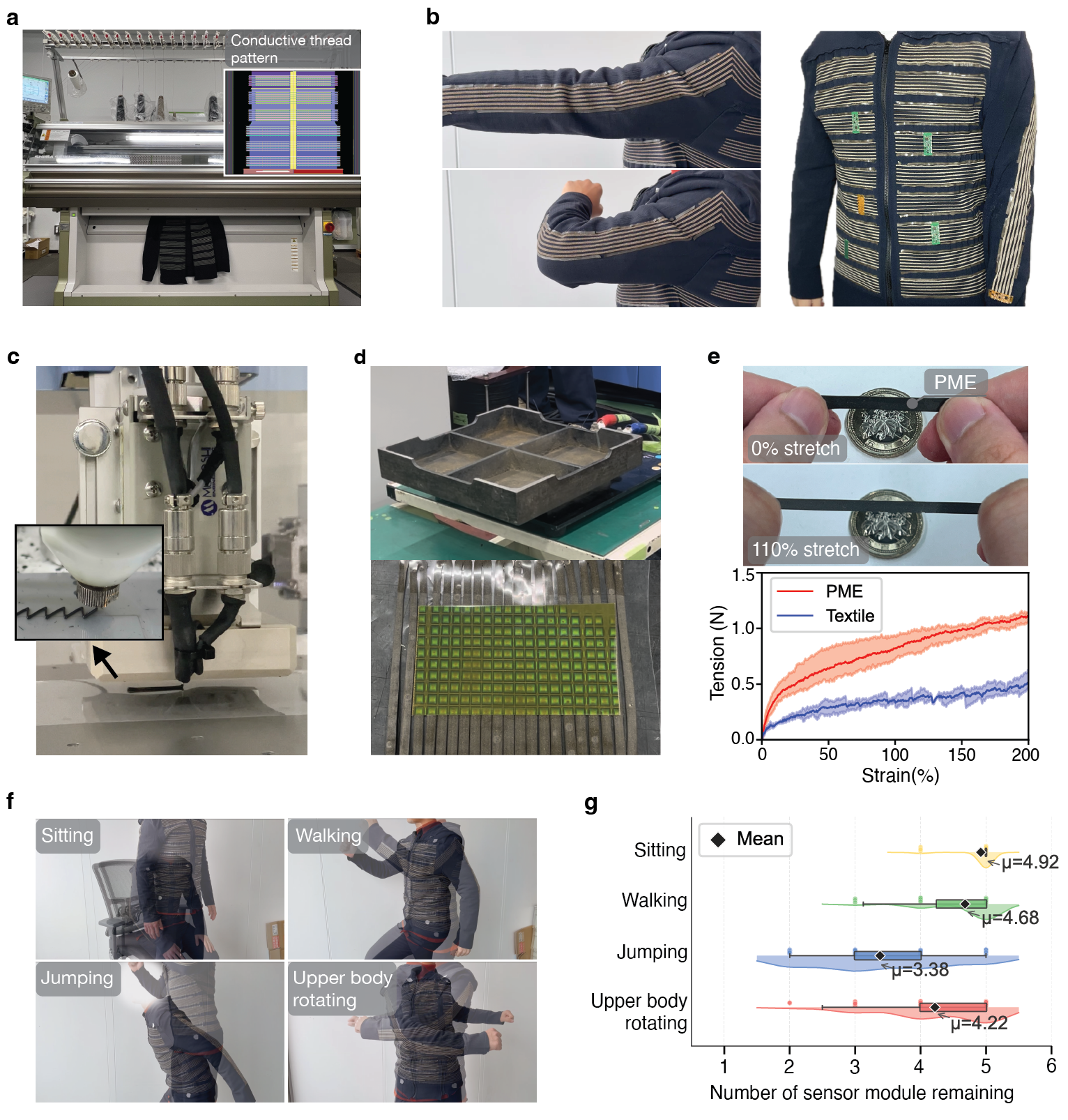}
  \caption{(a) The plug-n-play e-knit textile is designed in computer and fabricated by knitting machine, (b) while maintains good stretchability and large-scale coverage. The fabrication process of PME, including (c) printing the PME strip, (d) magnetisation and obtaining the magnetic pole direction. (e) Relationship between the tension and the strain of the wide PME. The solid line denotes the mean and the shaded areas depict the maximum deviation of the three PME testing samples. (f)(g) The evaluation process and result of PME's connection stability while sitting, walking, jumping and upper body rotating.}
  \label{fig:PME}
\end{figure*}

To enable robust sensor attachment without causing mechanical damage to the textile, we utilize PME for its inherent stretchability and stable connection.
The fabrication procedure of PME is similar to~\cite{Shembekar_PME_2021}, as shown in~\autoref{fig:PME}c and d, we mix a polyurethane resin with a neodymium powder. 
The mixed polyurethane resin is extruded into meander strips via an inkjet printer and kept for solidification. 
Then, the solidified material is placed in a magnetizing power supply machine (ES-4030-30, Denshijiki Industry Co., Ltd.) and applied to an ultra-strong magnetic field of 6 tesla. 
The magnetized PME will feature magnetic poles appearing alternately at 5mm intervals. We use adhesive to attach the PME at the edge of each wiring group, so that the flexible PCBs equipped with four rigid magnets can be strongly connected with the wiring. 
We test the stretchability of PME in the Tensile and Compression Testing Machine (MCT-2150W, A\&D Company Ltd.) with three \SI{40}{\mm} $\times$ \SI{5}{\mm} sized PME samples. 
\autoref{fig:PME}e shows that PME provides a compatible stretchability to textile, which allows natural body movement for daily wear.

\begin{figure*}[h!]
  \centering
  \includegraphics[width=1.0\textwidth]{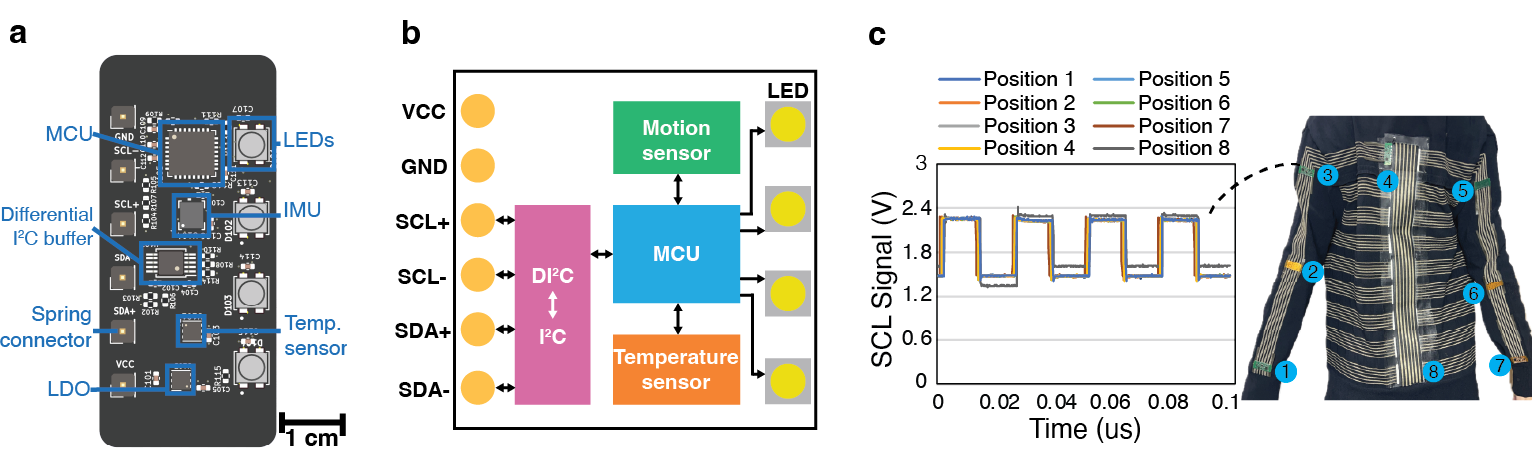}
  \caption{Design overview of the sensing module, including a low dropout regulator(LDO), temperature sensor, differential I$^2$C buffer, inertial measurement unit(IMU), LEDs, MCU and spring connector.}
  \label{fig:board}
\end{figure*}

\begin{figure*}[h!]
  \centering
  \includegraphics[width=1\textwidth]{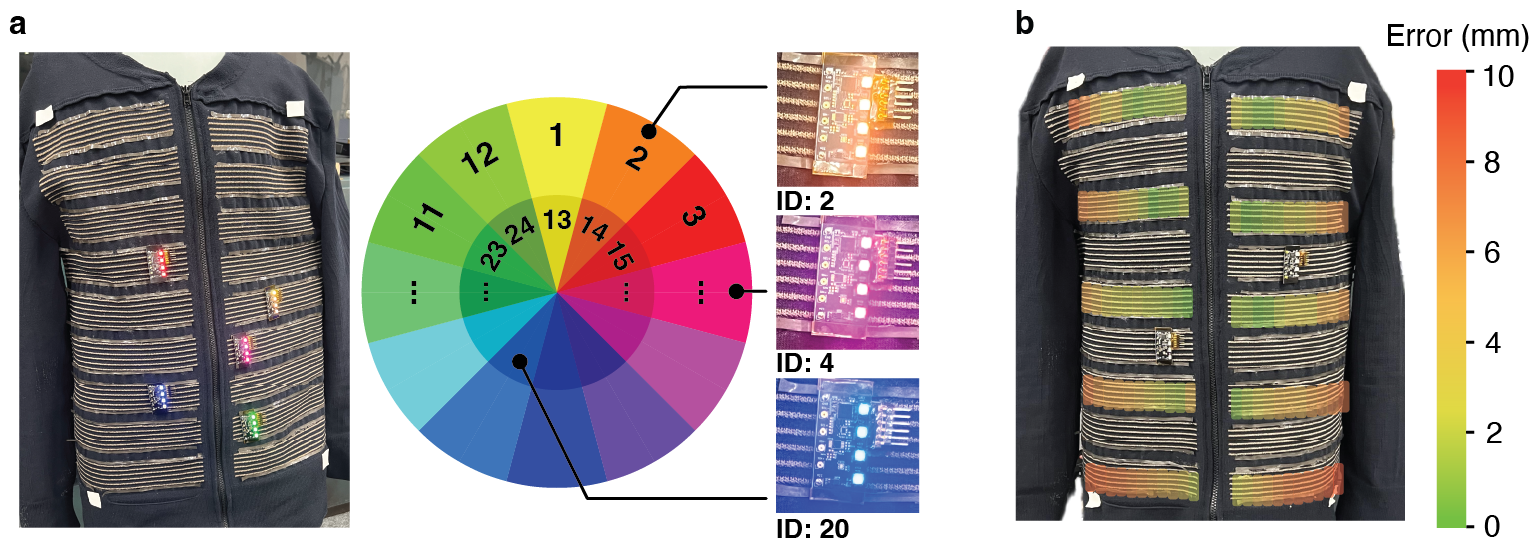}
  \caption{(a) Plug-n-play e-knit can recognize 24 LED colors based on the difference of hue and brightness. (b) The LED localization error is tested on five groups of conductive threads.}
  \label{fig:LED}
\end{figure*}

The e-textile typically functions during the user's movement, so the soft magnet connection in the plug-n-play e-knit needs to be robust against user motion~\cite{tseghai2020review}. 
We evaluated the robustness by monitoring the communication status of the five sensing modules during four types of user motions including walking, running, jumping, and upper body rotating (\autoref{fig:PME}f).
Note that the sensing modules were placed at the left wrist, right wrist, back, chest, and waist, and the motion period was 5 seconds.
\autoref{fig:PME}g shows the remaining number of the sensing modules for each motion test 50 times. 
This is due to the additional transient forces induced by the fabric's deformation and local 'whipping effect,' the sensor module requires a minimum retention force of \SI{0.6}{\N} to maintain a stable connection.
The wider PME design could solve this issue by increasing the surface magnetic force of the PME.

\subsection{Sensing Module}

To implement various functions on e-textiles while maintaining wear comfort, we need lightweight and customizable sensing modules~\cite{casciani2023unpacking}. 
The energy and signal of the plug-n-play e-knit can be transmitted throughout the entire garment, providing the conditions for this. Here, we introduce the IMU and temperature sensing module as an example. 
As shown in~\autoref{fig:board}a and b, the sensing module consists of flexible PCB with a low-power MCU (STM32L432KC, STMicroelectronics), Low Dropout(LDO) (TLV76733, Texas Instruments), IMU sensor~(ICM20948, TDK InvenSense), temperature sensor~(TMP117, Texas Instruments), Differential I$^2$C buffer chip (PCA9615, NXP Semiconductors), LED~(5050RGB-SK6218, XINGLIGHT) and spring connectors which align with the conductive threads. 
We employ adhesive to attach magnets at the edge of the sensing modules for the magnetic connection with the PME. Each module receives power from and communicates with the hub via the differential I$^2$C protocol. 
The use of a lightweight flexible PCB allows for stable recording of the 9-axis IMU motion data and real-time temperature data. The module is \SI{40}{mm}$\times$\SI{15}{mm} in size and weighs \SI{1}{g}.

\begin{figure*}[h!]
  \centering
  \includegraphics[width=1\textwidth]{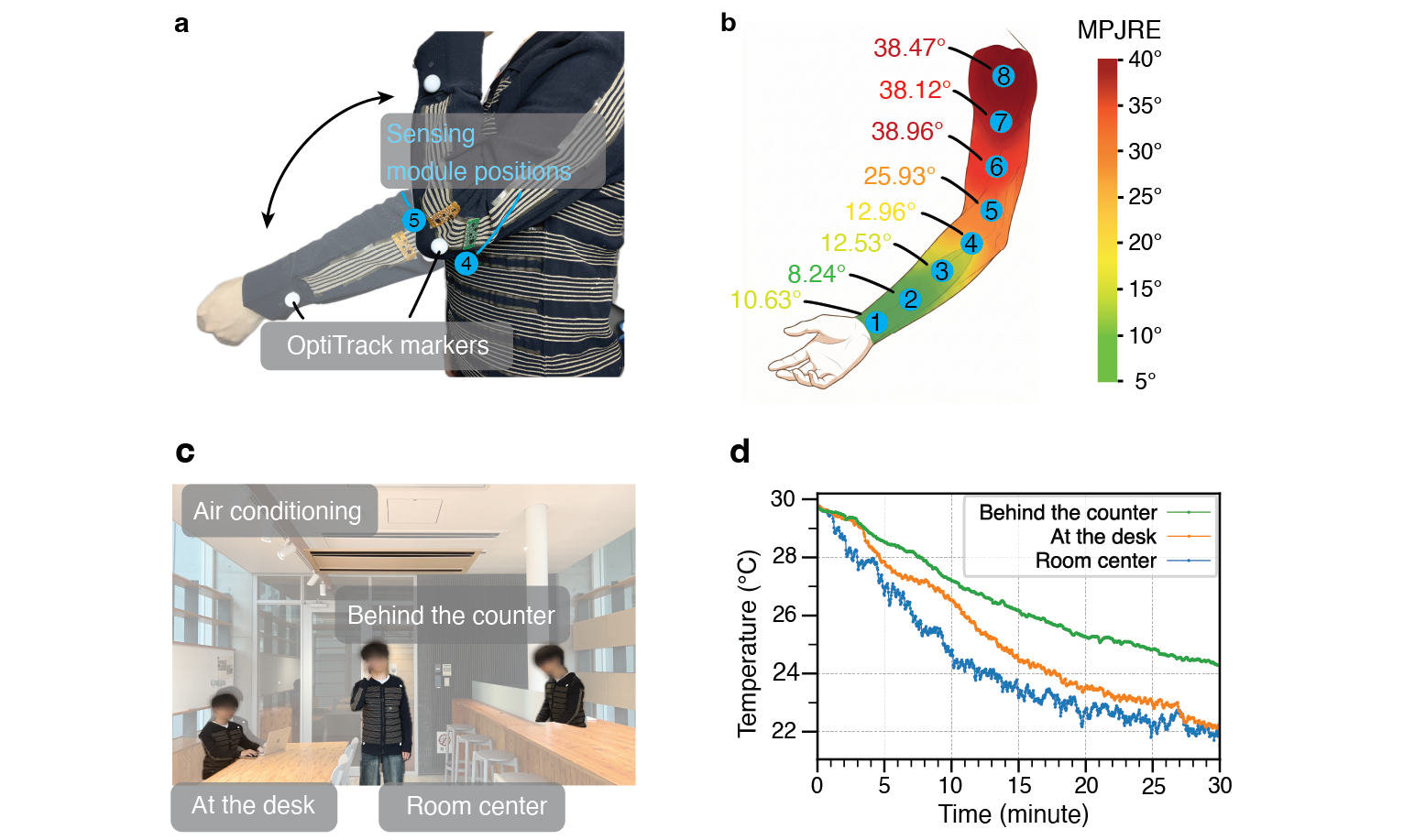}
  \caption{Application example: (a) The sensing module is positioned at eight points between the wrist and shoulder to actively test the optimal placement for capturing forearm movement, and (b) get Mean Per Joint Rotation Error (MPJRE) data for each position. And (c)(d) testing temperature changes at the desk, in the center of the room, and behind the counter.}
  \label{fig:Application}
\end{figure*}
For large-scale e-textile prototyping, the maximum communication range between the sensing modules and hubs must be over \SI{1.5}{\m} (i.e., the distance between the left wrist and hem of the garment)~\cite{zhu2022scaling}.
However, the signal attenuation in the long conductive wire causes communication failure~\cite{moradi2018signal}.
Therefore, we tested the performance of the two longest channels in the plug-n-play e-knit.
We chose conductive paths between the ends of the two sleeves and one sleeve to the corner of the garment.
We placed one sensing module, which serves as the data transmitter, on the right wrist (Position 1). 
Then, we tested the signal attenuation at the seven positions (Positions 2-8).
The voltage of the differential  I$^2$C signal (i.e., SCL signal) at the seven positions and the transmitting signal from the data transmitter are shown in~\autoref{fig:board}c. 
Although the signalat the edge (Position 8) is distorted, the differential I$^2$C receiver reliably decodes the distorted signal by making the voltage difference between the paired lines.
As a result, plug-n-play e-knit enables the differential I$^2$C communication without the signal loss.


\subsection{Sensing Module Localization}

For rapid prototyping, the system must know exactly where each sensor is located~\cite{noda_I2We_2019}.  We use an external camera and built-in LEDs to automatically identify and locate each sensor.
The system operates in two steps. 
First, it identifies the module’s ID by its specific color properties: hue and brightness (value). 
Second, it calculates the physical coordinates of the module relative to four reference markers placed at the corners of the textile. By using these markers, the system can compensate for the garment's orientation.

We first evaluated the maximum number of modules the system can distinguish. By testing different color ranges, we found that the camera requires a minimum hue difference of \qty{30}{\degree} to reliably tell two modules apart (\autoref{fig:LED}a). 
This provides \num{12} unique color IDs. 
While saturation was too sensitive to ambient light, we successfully used pulse-width modulation (PWM) to create \num{2} distinct brightness levels. 
Combined, the system supports up to \num{24} unique sensor IDs (\num{12} hues $\times$ \num{2} levels).
Next, we quantified the positional accuracy. 
We compared the coordinates calculated by the camera with ground-truth measurements taken by hand.
We moved a module with a red LED across various points on the textile and recorded the error. 
As shown in \autoref{fig:LED}b, the error remained below \qty{5}{\milli\meter} for most of the garment's surface. 
Near the corners, the maximum error reached \qty{9}{\milli\meter}, which we attribute to minor lens distortion from the wide-angle camera. 
This level of precision is more than sufficient for applications like body temperature mapping and joint tracking, where sensor placement is important but millimeter-level shifts do not alter the overall physiological data.


\section{Application Examples}
\subsection{Motion Capturing for Finding the Optimal Sensor Placement}

Finding the ideal sensor location is critical for accurate motion tracking. 
As proof of concept, we used plug-n-play e-knit to identify the optimal motion sensor position for capturing forearm flexion. 
We tested eight different positions along the arm, from the wrist to the shoulder, by simply snapping the IMU modules onto the garment (\autoref{fig:Application}a).
Note that the ground-truth data was measured by OptiTrack.
The experiment revealed that the most accurate data was obtained just below the wrist, with a Mean Per Joint Rotation Error (MPJRE) of \qty{8.24}{\degree}. Surprisingly, placing the sensor directly on the wrist was less accurate (\qty{10.63}{\degree}), as unconscious wrist twisting during forearm movement introduced noise into the signal. Tracking accuracy dropped significantly once the sensor was moved past the elbow joint (\autoref{fig:Application}b).
This experiment highlights the efficiency of our prototyping platform. In traditional e-textiles, testing eight different locations would require labor-intensive and destructive rewiring. 
With plug-n-play e-knit, the user simply plugs and unplugs the modules to find the best configuration in minutes. Furthermore, our system supports simultaneous multi-node testing; we successfully recorded data from above and below the elbow at the same time. This capability allows researchers to evaluate entire sensor networks rapidly, ensuring the best possible performance for wearable robotic interfaces.

\subsection{On-textile Temperature Mapping in Room}

To design efficient air conditioning, we must understand how cooling varies across a room~\cite{shinoda2021differences}. We used the plug-n-play e-knit to monitor these temperature changes in real-time. We tested the system for \qty{30}{\minute} in a \qty{30}{\degreeCelsius} room, with the air conditioner (SZRC50BYV, Daikin) set to \qty{20}{\degreeCelsius} at maximum fan speed.
During the test, a user wearing the e-knit moved between three locations: the room center, a desk, and a kitchen counter (\autoref{fig:Application}c). Our data revealed distinct cooling curves for each spot. The area near the desk cooled most slowly because the furniture blocked the airflow and the user’s seated posture created a pocket of warm air (\autoref{fig:Application}d). In contrast, the center of the room showed the largest temperature fluctuations, as the air conditioner's swing mode blew cold air directly toward that spot.
The experiment confirms two key strengths of our platform. First, it captured precise, position-aware data that would be missed by a single wall-mounted sensor. Second, the user performed all tasks naturally without feeling restricted. The stable, continuous data stream proves that plug-n-play e-knit is a reliable tool for long-term physiological and environmental monitoring in daily life.

\section{CONCLUSION}

This paper proposes plug-n-play e-knit, a platform that transforms large-scale garments into reconfigurable, non-invasive prototyping tools. 
By combining industrial machine knitting with soft magnetic connectors, we allow users to freely and repeatedly rearrange sensor modules without damaging the fabric. 
Our experiments in motion tracking and temperature mapping demonstrate that this system can capture high-quality data across the entire body, paving the way for the ubiquitous development of smart clothing.
Despite these promising results, several technical challenges remain. First, because the conductive channels are currently exposed to ensure a good connection, the fabric can short-circuit when compressed, particularly in high-flexion areas like the inner arm.
Second, while our magnetic connectors are stable during low-intensity activities, they can detach during vigorous movements such as jumping. There is a fundamental trade-off here: increasing the magnetic force improves the connection but also increases the stiffness of the garment. 
Future iterations will investigate more flexible magnetic elastomer formulations and additional support strips to balance security with comfort.
Finally, although the DI2C-based communication protocol can support up to 127 modules, the color-based identification method is limited to recognizing only 24 modules due to factors such as lighting conditions and the LED bulb color range. In the future, we will encode the flashing frequency and pattern of four LEDs as an identity ID to achieve greater identification capacity.

\addtolength{\textheight}{-3cm}   



\bibliographystyle{IEEEtran}
\bibliography{reference}

\end{document}